\documentclass[fleqn,10pt]{wlscirep}
\usepackage[utf8]{inputenc}
\usepackage[T1]{fontenc}

\usepackage{graphicx}   
\usepackage{xcolor}    
\usepackage{caption}   
\usepackage{cite}
\usepackage{lineno}     

\usepackage{amsmath}
\usepackage{tabularx,booktabs,multirow}

\title{110 GHz, 110 mW Hybrid Silicon-Lithium Niobate Mach-Zehnder Modulator}
\author[1,*]{Forrest Valdez}
\author[1]{Viphretuo Mere}
\author[1]{Xiaoxi Wang}
\author[2]{Nicholas Boynton}
\author[2]{Thomas A. Friedmann}
\author[2]{Shawn Arterburn}
\author[2]{Christina Dallo}
\author[2]{Andrew T. Pomerene}
\author[2]{Andrew L. Starbuck}
\author[2]{Douglas C. Trotter}
\author[2]{Anthony L. Lentine}
\author[1,**]{Shayan Mookherjea}

\affil[1]{University of California, San Diego, Department of Electrical and Computer Engineering, 9500 Gilman Drive, MC 0407, La Jolla, California, USA}
\affil[2]{Sandia National Laboratories, Applied Microphotonic Systems, Albuquerque, New Mexico 87185, USA}
\affil[*]{fgvaldez@eng.ucsd.edu}
\affil[**]{smookherjea@ucsd.edu}

\begin{abstract}
High bandwidth, low voltage electro-optic modulators with high optical power handling capability are important for improving the performance of analog optical communications and RF photonic links. Here we designed and fabricated a thin-film lithium niobate (LN) Mach-Zehnder modulator (MZM) which can handle high optical power of 110 mW, while having 3-dB bandwidth greater than 110 GHz at 1550 nm. The design does not require etching of thin-film LN, and uses hybrid optical modes formed by bonding LN to planarized silicon photonic waveguide circuits. A high optical power handling capability in the MZM was achieved by carefully tapering the underlying Si waveguide to reduce the impact of optically-generated carriers, while retaining a high modulation efficiency. The MZM has a $V_\pi L$ product of 3.1 V.cm and an on-chip optical insertion loss of \textcolor{black}{1.8} dB.
\end{abstract}
\begin{document}
\flushbottom
\maketitle
\thispagestyle{empty}
\section*{Introduction}
Electro-optic modulators (EOMs) are essential components for optical communications, radio frequency (RF)-photonic links, frequency combs, optical phased arrays and optical information processing. In recent years, thin-film lithium niobate (TFLN) EOMs with a high electro-optic 3-dB bandwidth of at least 100 GHz have been reported with a low voltage-times-length ($V_\pi L$) product and low optical insertion loss\cite{wang2018integrated,Wang2019,yang2022monolithic,xu2022dual}. However, these demonstrations have been performed at relatively low optical power levels less than 10 mW. The ability to handle high optical powers without increasing $V_\pi$ would be beneficial, by increasing the RF link gain and reducing the noise figure of an analog communications system\cite{Ackerman2005,Cox2006a,williamson2008rf}. For a given photodiode responsivity, reducing $V_\pi$ and increasing the optical power are both important in the noise figure of an RF photonic link. Recently, a higher optical power level, around 25 mW, was shown using TFLN Mach-Zehnder modulator (MZM) with about 50 GHz bandwidth \cite{shams2022electrically}. Here, we report the first TFLN MZM which achieves both above-100 GHz bandwidth and above-100 mW optical power handling capability, while also achieving a low $V_\pi L$ product and low on-chip optical insertion loss, and integration with silicon photonics. 

TFLN-based waveguides have been made using various methods such as etching, blade-dicing or polishing of ridge or rib waveguides, and rib-loading of a TFLN slab using other materials \cite{rabiei2005lithium,hu2007lithium,wang2018integrated,zhao2020shallow,courjal2011high,wu2018long,liang2021monolithically,he2019high,rao2015heterogeneous,weigel2018e,Ahmed2020,mere2022modular}. Our approach is based on hybrid optical modes, in which the waveguide core consists of a combination of silicon (Si) and LN, and the cladding consists of silicon dioxide ($\mathrm{SiO}_2$). This approach takes advantage of both the scalability of Si photonics, which is fabricated on large wafers using mature foundry processing, and the EO properties of crystalline LN without having to etch or pattern it. The Si waveguides are formed in the crystalline layer of a silicon-on-insulator (SOI) wafer, and are tapered in the hybrid region to implement inter-layer optical mode transitions; the TFLN layer is not etched or patterned. The Si waveguides can be continuously connected, without breaks, to other photonic components outside the bonded region, and thus present a simple and scalable way to include ultrahigh bandwidth EOMs with other photonic components\cite{weigel2016,mere2022modular}.  

At higher optical powers, Si waveguides suffer from two-photon absorption (TPA) and free carrier absorption (FCA) and dispersion (FCD) effects \cite{tsang2002optical,bristow2007two,lin2007nonlinear,leuthold2010nonlinear}. The nonlinear optical absorption of amorphous Si is sensitive to processing \cite{wathen2014non} and can even exceed that of crystalline Si by a factor of 30 or more \cite{ikeda2007enhanced}. Therefore, it is preferable in hybrid TFLN design to use a layer of crystalline Si waveguides rather than depositing Si on top of TFLN which has been a popular strategy \cite{cao2014hybrid,wang2017amorphous}. Also, by carefully designing the Si layer dimensions and tapers, the amount of optical power in the Si region can be minimized while still remaining in the single-mode regime, which is necessary for a high-bandwidth MZM. While our strategy does not totally eliminate the nonlinear loss of Si, it can keep the impairments of the hybrid mode adequately low for practically-useful optical power levels over the device lengths needed for a high-bandwidth, low-voltage MZM.  

\begin{figure}[ht!]
\includegraphics[width=17.5cm]{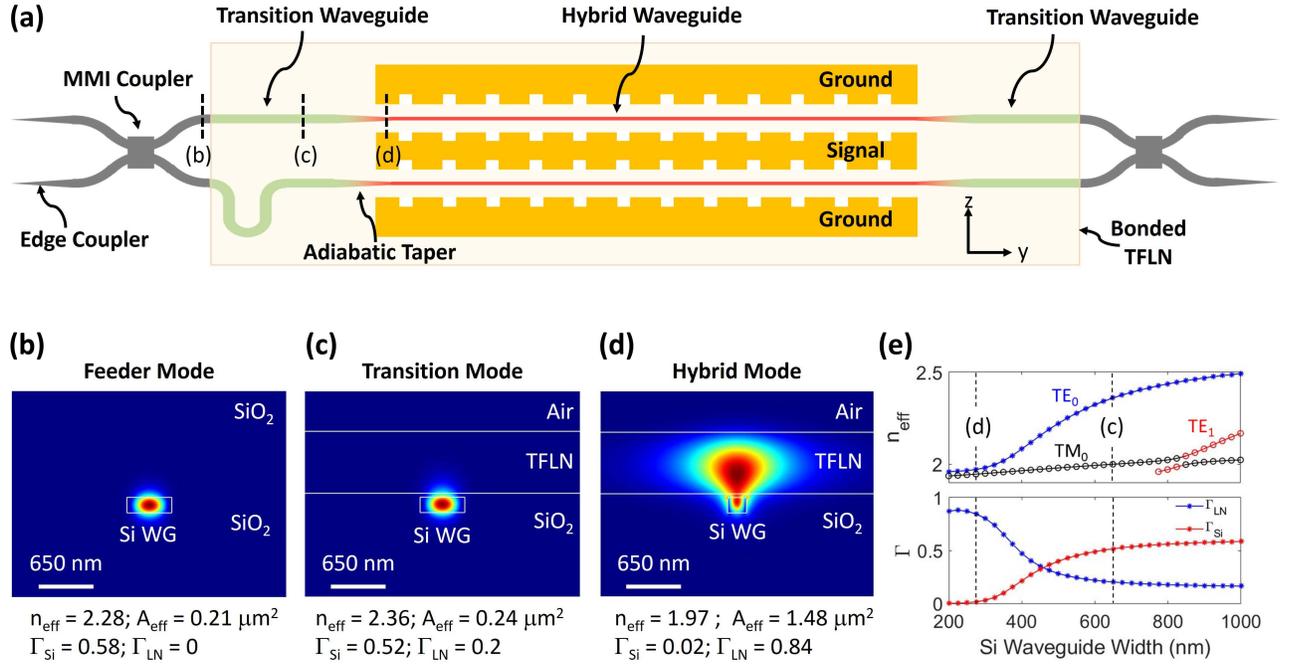}

\caption{\textbf{(a)} Top-view schematic (not to scale) of the Si/LN MZM. The transition region consists of the transition waveguide and adiabatic tapers (shown in green), and the phase shifter section consists of the hybrid waveguide (shown in red). Cross-sectional simulated Poynting vector of the \textcolor{black}{\textbf{(b)} feeder Si waveguide (no LN film),} \textbf{(c)} transition Si waveguide, and \textbf{(d)} hybrid Si/LN waveguide. \textbf{(e)} The simulated effective index of the first three Si/LN modes (top), and the confinement factor in the TFLN layer and Si waveguide (bottom) as a function of Si waveguide width at $\lambda = 1550$ nm. The dashed lines indicate the modes shown in panels \textbf{(c)} and \textbf{(d)}.}
\label{fig1}
\end{figure}

\section*{Results}
\subsection*{Theory and Design}
The phase shifter sections of the MZM, shown by the regions that are colored red in Fig. \ref{fig1}(a), consist of a long single-mode waveguide in which the transverse localization of the hybrid optical mode is controlled by narrowed Si features (275 nm wide and 150 nm tall). \textcolor{black}{Adiabatically tapering down the Si waveguide reduces the effective index of the guided mode ($n_\mathrm{eff}$), and transitions the confinement from the high index Si to the LN layer, as shown in Fig. \ref{fig1}(e).} For this TE-polarized hybrid mode, light exists mainly in the LN section, and as shown in Fig. \ref{fig1}(d), only 2\% of the Poynting vector resides in Si, with 84\% in TFLN, and the rest in the cladding. Therefore, the nonlinear impairments can be expected to be much lower than for a conventional Si carrier-depletion MZM where most of the light is in Si \cite{witzens2018high}. Because the amount of light in Si is so small, the hybrid TFLN phase shifter does not have a significant nonlinear optical penalty compared to a etched TFLN design.

The transition waveguide regions, shown by the regions that are colored green in Fig. \ref{fig1}(a), consist of wider Si features (650 nm wide and 150 nm tall) which confines light mainly in Si (Fig. \ref{fig1}(c): 52\% of the Poynting vector magnitude of the fundamental waveguide mode resides in Si, only 20\% in the thin film of LN, and the rest in the cladding). \textcolor{black}{Figure 1(b) and 1(c) show the simulated mode profile of the Si waveguides without and with the LN film above, respectively. The effective index ($n_\mathrm{eff}$) difference between these modes is $0.08$ which results in an estimated loss of only about 0.074 dB. Because the mode fractions in Si ($\Gamma_\mathrm{Si}$) are similar, we do not expect this loss to significantly change at higher powers when the refractive index of Si changes slightly. Overall, this loss should be negligible compared to other losses in the device that are described below in detail.} This allows light to propagate with minimal loss ($<$ 0.1 dB) across the bonded LN edge, where one side has LN as part of the multi-layer stack and the other side does not. We can thus mitigate sensitivity to micro-roughness in the diced edges of the LN chip, which simplifies the fabrication process. The length of the transition waveguide segment is short: a few tens of microns can be sufficient to traverse the edge of the LN film, and thus, the length-integrated effects of these nonlinear impairments are low as shown using numerical simulations which are discussed below. 

Silicon waveguides in which the mode closely resembles the transition mode (but with no TFLN layer on top, as shown in Fig. \ref{fig1}(b)) were used to form bends and the 3-dB multi-mode interference (MMI) couplers. These segments can also be kept very short, and the length-integrated nonlinear impairments are small at the relevant optical power levels. Our test chip also include tapered-waveguide edge couplers for guiding light from optical fibers to the waveguides. In our experiments, it was these segments, rather than the waveguides and transitions that constituted the MZM, were damaged at the highest optical power levels (above 150 mW). A detailed study of high-power fiber-to-waveguide coupling is beyond the scope of this report.

Nonlinear impairments in a traveling-wave MZM device can, in principle, limit the intrinsic MZM EO bandwidth, insertion loss (IL), and half-wave voltage ($V_\pi$). For a given material dielectric stack, high speed EO modulation depends on three design rules: 1) matching the velocities of the RF and optical waves; 2) matching the transmission line, source, and load impedances (typically 50 $\Omega$); and 3) minimizing the RF propagation losses \cite{ghione2009semiconductor,noguchi1995,Cox1997,Honardoost2018b}. The density of photo-generated electronic carriers caused by TPA is 
\begin{equation}
N_\mathrm{TPA}(P) = G_\mathrm{TPA}\tau_e = \frac{\beta\tau_e}{2h f} \left( \frac{P}{A_\mathrm{eff}}\right)^2,
\label{eqnNTPA}
\end{equation}
where $\beta$ is the TPA coefficient of Si (assumed to be 1.0 cm/GW \cite{bristow2007two}), $h f$ is the energy of the photon, $P$ is the optical power, and $A_\mathrm{eff}$ is the effective area of the optical mode.  The electronic carrier lifetime $\tau_e$ in crystalline Si waveguides is typically in the range of 1 to 100 ns, and depends on the waveguide geometry and the surface properties \cite{kuwayama2003interface}. We estimate our 150 nm thick strip waveguide to have an upper limit lifetime of 15~ns \cite{claps2004influence}. These photogenerated carriers lead to additional optical losses and a shift in refractive index in the Si waveguide via the plasma dispersion effect \cite{soref1987electrooptical} given by, 
\begin{subequations}
\begin{equation}
\alpha_\mathrm{FC}(P) = 1.45\times10^{-17}N_\mathrm{TPA}
\end{equation}
\begin{equation}
\Delta n_\mathrm{FC}(P) = -8.8\times10^{-22}N_\mathrm{TPA} - 8.5\times10^{-18} {N_\mathrm{TPA}}^{0.8}.
\end{equation}
\end{subequations}

If the change in the modal cross-sectional area $A_\mathrm{eff}$ is negligible, the evolution of the guided optical power as a function of propagation distance ($y$) with linear and nonlinear losses is
\begin{equation} \label{eq_IL}
\frac{dP(y)}{dy} = -\left(\alpha_l+\alpha_\mathrm{FC} \Gamma_\mathrm{Si}\right) P - \frac{\beta }{A_\mathrm{eff}}\Gamma_\mathrm{Si} P^2 
\end{equation}
where $\alpha_l$ is the linear absorption loss of the waveguide and $\Gamma_\mathrm{Si} = \displaystyle \left(\iint_\mathrm{Si}{|E_\mathrm{o}|^2\mathrm{dA}}\right)\cdot\left( \iint_{-\infty}^{+\infty}|E_\mathrm{o}|^2\mathrm{dA}\right)^{-1}$ is the optical confinement factor in the Si waveguide in terms of the electric field distribution of the optical mode, $E_\mathrm{o}(x,y)$.

\begin{figure}[ht!]\centering

   \includegraphics[width=\textwidth]{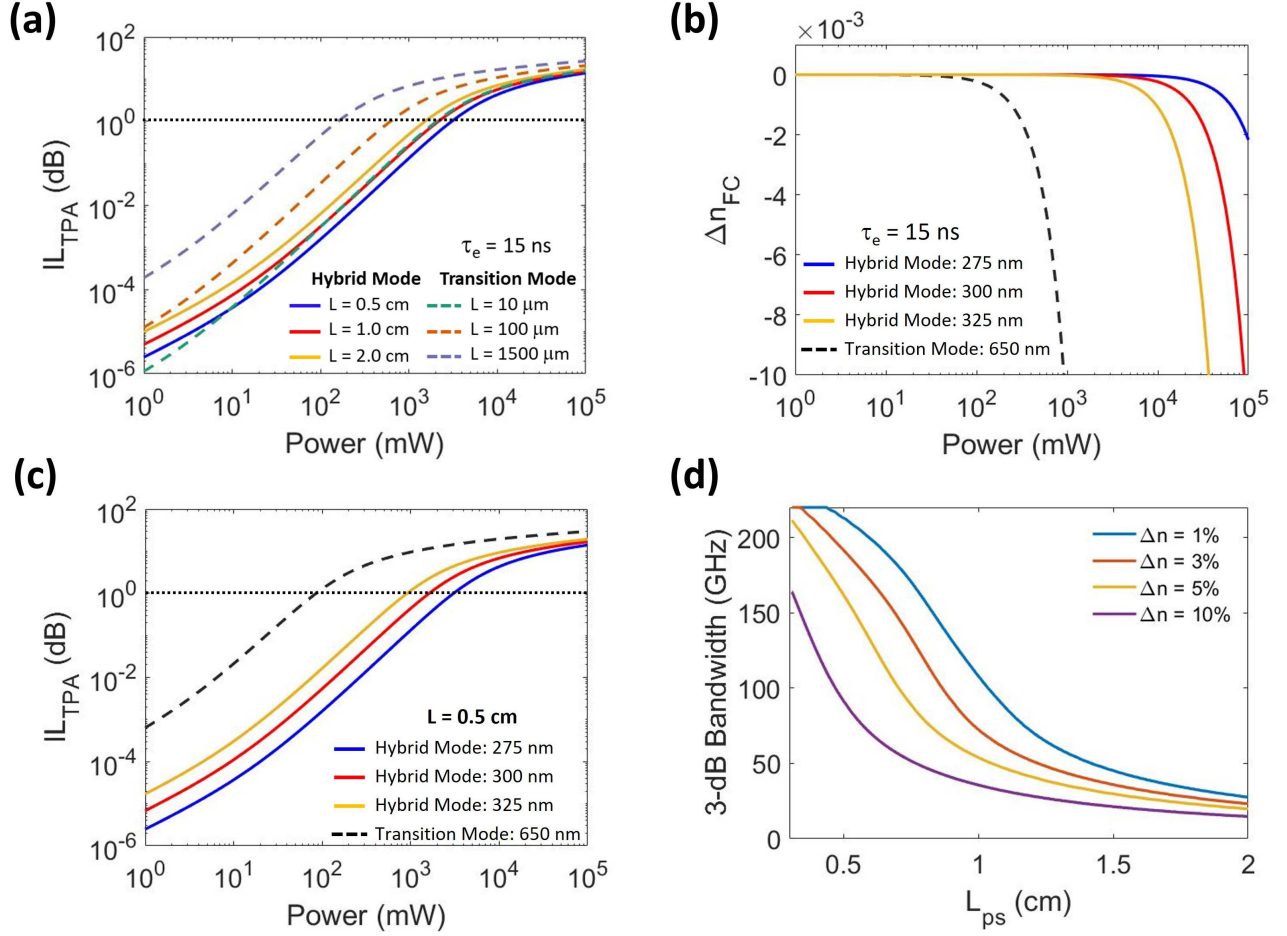}

   \caption {\textbf{(a)} The calculated insertion loss (IL) due to TPA as a function of peak optical power in the hybrid waveguide mode (solid) and the transition mode (dashed) assuming different lengths of the phase-shift segment. \textbf{(b)} The change in refractive index due to the additional free carriers generated via TPA as a function of peak optical power for the hybrid and transition modes. \textcolor{black}{\textbf{(c)} The calculated IL due to TPA as a function of peak optical power in the hybrid (solid) and transition (dashed) mode assuming a length of 0.5 cm for different waveguide widths.} \textbf{(d)} The calculated 3-dB bandwidth of a traveling-wave MZM as a function of $L_\mathrm{ps}$ assuming impedance matching to a 50 $\Omega$ load, and for different values of the index mismatch, $\Delta n_\mathrm{mo}$. Abbreviations: WG: waveguide, TFLN: thin-film lithium niobate.}
\label{fig2}
\end{figure}

Figure \ref{fig2}(a) shows the calculated change in IL due to nonlinear effects (TPA and FCA) by numerically solving Eq. (\ref{eq_IL}) for different lengths of hybrid waveguide (solid lines) and transition waveguides (dashed lines). We assume that the electronic carriers generated by light are confined to the Si waveguide, via $\Gamma_\mathrm{Si}$ in Eq.~(\ref{eq_IL}). The cross-sectional modal areas $A_\mathrm{eff}$ of the transition and hybrid modes are calculated to be 0.24 $\mathrm{\mu m}^2$ and 1.48 $\mathrm{\mu m}^2$, respectively. For the maximum considered lengths of 2 cm and 0.15 cm for the hybrid and transition waveguides in Fig. \ref{fig2}(a), the 1 dB IL penalty due to TPA occurs at an optical power ten times higher in the hybrid waveguide than in the transition waveguide. If the transition section is reduced to 100~$\mathrm{\mu}$m (which is approximately the minimum for an adiabatic taper between the modes shown in Fig.~\ref{fig1}(c) and \ref{fig1}(d)\cite{fu2014efficient}), then the total additional IL should be less than 1 dB for optical powers up to 0.6 W. 

The electro-optic response (EOR) as a function of RF frequency and optical power is\cite{ghione2009semiconductor}
\begin{subequations}
\begin{equation} \label{eq_EOR}
m(\omega,P)=\frac{R_{L}+R_{G}}{R_{L}}\left|\frac{Z_\mathrm{in}}{Z_\mathrm{in}+Z_\mathrm{G}}\right|\left|\frac{(Z_\mathrm{L}+Z_\mathrm{c})F(u_{+}(P))+(Z_\mathrm{L}-Z_\mathrm{c})F(u_{-}(P))}{(Z_\mathrm{L}+Z_\mathrm{c})\exp[\gamma_\mathrm{m}L_\mathrm{ps}]+(Z_\mathrm{L}-Z_\mathrm{c})\exp[-\gamma_{m}L_\mathrm{ps}]}\right|,
\end{equation}
where the input impedance of the coplanar traveling wave electrodes is defined as
\begin{equation}
Z_\mathrm{in}=Z_\mathrm{c}\frac{Z_\mathrm{L}+Z_\mathrm{c}\mathrm{tanh}(\gamma_\mathrm{m}L_\mathrm{ps})}{Z_\mathrm{c}+Z_\mathrm{L}\mathrm{tanh}(\gamma_\mathrm{m}L_\mathrm{ps})},
\end{equation}
and the complex RF propagation constant is
\begin{equation} \label{eq_gm}
\gamma_\mathrm{m}=\alpha_\mathrm{m}+\frac{j\omega}{c}n_\mathrm{m}.
\end{equation}
Here, $R_\mathrm{L,G}$ are the load and generator resistances, $Z_\mathrm{L,G,c}$ are the load, generator, and characteristic impedances, $\alpha_\mathrm{m}$ is the RF propagation loss, $n_\mathrm{m}$ is the RF effective index, and $L_\mathrm{ps}$ is the phase-shifter interaction length.
$F\left[u_{\pm}(P)\right]$ and $u_{\pm}(P)$ are defined as
\begin{equation}
F\left[u_{\pm}(P)\right]=\frac{1-\exp\left[u_{\pm}(P)\right]}{u_{\pm}(P)}
\end{equation}
and
\begin{equation} \label{eq_upm}
u_{\pm}(P)=\pm\alpha_\mathrm{m}L_\mathrm{ps}+\frac{j\omega}{c}(\pm n_\mathrm{m}-n_\mathrm{g}(P))L_\mathrm{ps}.
\end{equation}
\end{subequations}
The optical power dependence in Eq.~(\ref{eq_EOR}) is included by the associated FCD term affecting the optical group index, $n_g$, from photogenerated carriers in Eq.~(\ref{eq_upm}). There will be a similar shift in $V_\pi L$ due to FCD, where the effective modulator length, $L$, may be limited by the additional nonlinear losses [Eq.~(\ref{eq_IL})] and, thus, impact $V_\pi$. Since the hybrid MZM uses an x-cut TFLN film and a TE-polarized fundamental waveguide mode, the effects of FCD on $V_\pi$ can be expressed as
\begin{equation} \label{eqn_VpiL}
V_\pi L_\mathrm{ps}(P) = \frac{n_\mathrm{eff}(P)\lambda_0 G}{2n_e^4 r_{33}\Gamma_\mathrm{mo}}
\end{equation}
where $n_\mathrm{eff}$ is the hybrid mode effective index, $\lambda_0$ is the free-space wavelength, $G$ is the signal-to-ground electrode gap spacing, $n_e$ is the extraordinary index of TFLN, $r_{33}$ is the electro-optic coefficient of TFLN, and $\Gamma_{mo}$ is the overlap integral between the RF and optical modes in the electro-optic region.

Figure \ref{fig2}(b) shows the change in refractive index. Changes in the refractive index will effect the index matching condition ($\Delta n = (1 - n_m/n_g)\times100\%$) between the traveling RF and optical waves. Since the hybrid mode is predominantly localized in the LN film, the effect is expected to be negligible for optical peak powers less than 10 W in theory. Although high intensity optical inputs can also result in an increase in local temperature (and an associated thermo-optic effect), the time scales of such thermal shifts is much longer compared to the modulation speeds \cite{ruske2003photorefractive} and could be compensated by the bias controller. Furthermore, the photorefractive effect of TFLN at a wavelength of 1550 nm results in refractive index changes on the order of $10^{-5}$, \cite{jiang2017fast} which would only change $\Delta n$ by 0.005\%. 

Assuming the impedance matching condition is met, the 3-dB bandwidth of a traveling wave hybrid TFLN MZM is plotted as a function of the length of the phase shifter segment for different values of $\Delta n$ in Fig. \ref{fig2}(d) using Eq. (\ref{eq_EOR}). For $L_\mathrm{ps} = 0.5$ cm, to reduce the bandwidth by a factor of two (i.e., 3~dB), the $\Delta n$ caused by nonlinear high power effects would need to be about 10\%, corresponding theoretically to a peak power greater than 100 W, which is much greater than we are able to test experimentally. A MZM device using a longer phase shifter of length $1$ cm, would decrease the bandwidth by 3 dB for $\Delta n = 5\%$.

In this work, the total length of the MZM is about 1.7 cm, which includes the edge couplers, MMIs, transition region ($L_\mathrm{Tr} = 0.14$ cm), and the hybrid region ($L_\mathrm{Hb} = 1.4$ cm). The 3-dB 2x2 MMI couplers are 38 µm long, 6 µm wide, and are designed for the C-band. Bezier S-bends are implemented to separate the MZM arms to allow appropriate space for a standard 55 µm wide electrode. In the hybrid bonded region, the transition waveguides are adiabatically tapered down from a width of 650 nm to a width of 275 nm over a length of 300 µm which pushes the optical mode up into TFLN. These transition sections were intentionally made longer than necessary to simplify our bonding process. \textcolor{black}{If the waveguide had 300 nm width (instead of 275 nm), then $\Gamma_\mathrm{Si}$ increases slightly from $2\%$ to $3.3\%$, which will increase the number of photogenerated carriers in Si. Figure \ref{fig2}(b) and \ref{fig2}(c) show the additional index change and IL as the Si width changes for a hybrid waveguide length of 0.5 cm. For an optical input power of 110 mW, the additional IL due to TPA increases from 0.002 dB to 0.006 dB as the hybrid waveguide Si width increases from the target value of 275 nm to 300 nm.}

\subsection*{Hybrid Si/LN Chip Fabrication}

The hybrid Si/LN EOM was made by direct hydrophilic bonding without using an intermediate adhesive layer, such as benzocyclobutene (BCB) \cite{poberaj2009ion}\textcolor{black}{, whose performance and stability at high optical powers is uncertain.} 
The Si waveguides were built on a 200 mm diameter SOI wafer with 150 nm thick Si, 3 µm thick silicon dioxide (SiO$_2$), and 725 um thick high resistivity Si handle. The Si waveguide features were fabricated using deep-UV lithography followed by reactive-ion-etching. SiO$_2$ was deposited on top of the Si waveguide features and then the surface was planarized using a CMP process to achieve a bondable surface, leaving a thin SiO$_2$ layer ($< 50$ nm). The LNOI wafers were procured commercially from NanoLN, Jinan Jingzheng Electronics Co., Ltd. The LNOI is an x-cut 600 nm thick LN film with 2 µm thick buried SiO$_2$ and 500 µm thick Si handle.
\textcolor{black}{There is typically some variation of LN film thickness and thin SiO$_2$ layer of the SOI across the wafers. Variations in the thickness of SiO$_2$ layer between the bonded Si waveguide and LN film will change the mode fraction ($\Gamma$) contained in each region, and will also change the modulation efficiency \cite{weigel2018e}. However, based on the outcomes of the CMP process, this effect should be relatively small. For less than 20 nm change in the SiO$_2$ thickness, the corresponding change in $\Gamma_\mathrm{Si}$ or $\Gamma_\mathrm{LN}$ remains within a few percent of the target values. For each reticle of the larger SOI wafer, the thickness of the layer stack was verified by ellipsometry before bonding.} Based on the ellipsometric measurements of both the LN wafer and the processed silicon wafer after CMP, a die dicing plan is developed, and specific pairs of diced LN and Si chips are selected for bonding. In this way, the yield of devices from a wafer is increased, and the EO properties of each hybrid modulator will be closer to optimal.

\begin{figure}[ht!]
\includegraphics[width=\textwidth]{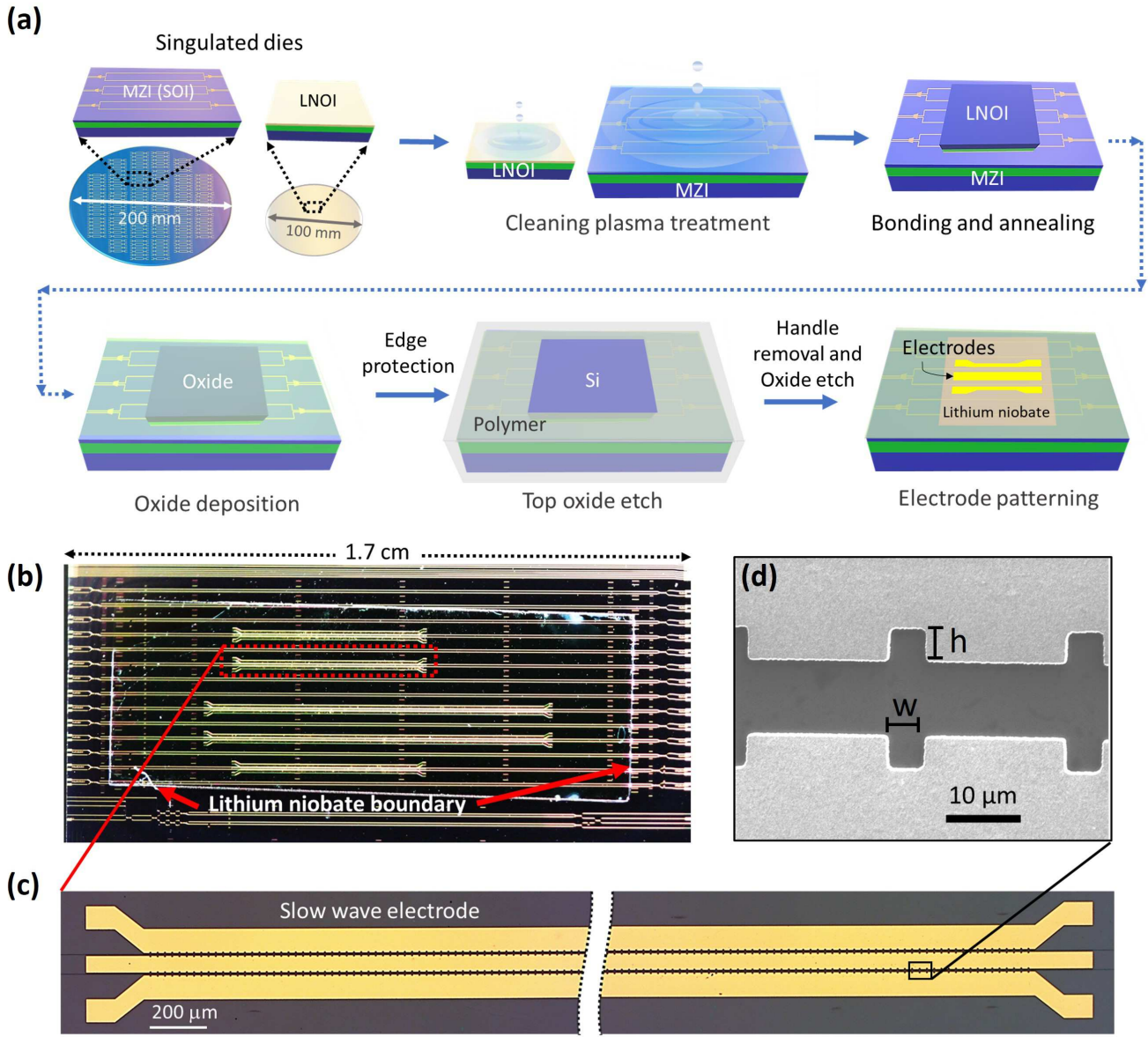}
\caption{\textbf{(a)} Fabrication process flow of the hybrid bonded Si/LN MZM. \textbf{(b)} Optical microscope image of hybrid bonded Si/LN chip with gold SWE. \textbf{(b)} The fabricated hybrid bonded Si/LN MZM chip. \textbf{(c)} A stitched optical microscope image of a 5 mm long SWE. \textbf{(d)} A scanning electron microscope image of the inductively loaded slot features.}
\label{figfab}
\end{figure}

Figure \ref{figfab}(a) illustrates the direct bonding hybrid Si/LN modulator fabrication process. Before bonding, the surfaces of the LNOI and SOI singulated-dies were cleaned using an RCA-1 process followed by plasma surface activation \cite{mere2022modular}. After plasma surface activation, the dies were soaked in deionized water. After drying the dies, the bonding was performed by contacting the two dies at room temperature. To improve the bond strength, the bonded sample was annealed using temperature cycles up to 300 $^\circ$C under an applied pressure of 9.8 N.cm$^{-2}$. At this stage, as described in our earlier work \cite{mere2022modular}, the bonded samples can be stored for processing at a later stage, if required. A plasma-enhanced-chemical-vapor-deposition (PECVD) was used to deposit a few microns of thick SiO$_2$ as the top cladding. Before the handle removal step, a polymer coating was applied to protect bonded chips except on top of the LNOI Si handle. Next, the oxide on top of the LNOI Si handle was etched using hydrofluoric acid. The LNOI Si handle was removed using a selective XeF$_2$-based etching process and then the oxide layer above LN was etched using HF. The polymer was cleaned using a solvent cleaning step, and a direct-laser writer was used to define the traveling wave electrode patterns using a negative photoresist. Finally, titanium and gold of thicknesses 20 nm and 750 nm, respectively, was deposited followed by a lift-off process to complete the fabrication of the electrodes. All steps from bonding of the TFLN chips to the final electrode fabrication were performed either at room temperature or at a modestly-elevated temperatures, not exceeding about 300 $^\circ$C.

For this device, an inductively loaded slow-wave electrode (SWE) structure [Fig. \ref{figfab}(b)-(d)] was designed to achieve velocity matching to the hybrid Si/LN optical mode. The gap between the signal and ground electrodes is 9 µm, and the signal width is 55 µm. The inductive loading feature is a periodic slot that is 5 µm wide (w) and 4 µm deep (h) with a period of 25 µm, as shown in Fig. \ref{figfab}(d). These periodic features increase the effective index of the microwave mode and slow the RF traveling wave to match the optical wave \cite{spickermann1993millimetre,sakashita1995preparation,shin2010conductor,Rosa2018,kharel2021breaking}. \textcolor{black}{Capacitively-loaded (T-rail based) SWE structures have been demonstrated in other reports \cite{kharel2021breaking,wang2022silicon} to achieve RF-optical velocity matching when using lower index substrates such as quartz or selectively-removed Si.} In this case, the hybrid Si/LN group index is simulated to be 2.32 and the slow-wave RF index is 2.34 at 110 GHz, resulting in $\Delta n < 1\%$ between the optical and RF waves before the nonlinear change of the index is taken into account. \textcolor{black}{For the fabricated structures with about 40 nm of CMP SiO$_2$ thickness, we estimate the peak electric field strength at 110 mW optical power in the oxide region of Fig. \ref{fig1}(c) and Fig. \ref{fig1}(d) to be about $5.4\times 10^4$ V.cm$^{-1}$ and $1.3\times 10^4$ V.cm$^{-1}$, which is smaller than the estimated breakdown field strength of the SiO$_2$, about $5\times 10^6 $ V.cm$^{-1}$.}

\subsection*{High-Speed, High-Power Measurements} 

\begin{figure}[ht!]
\includegraphics[width=\textwidth]{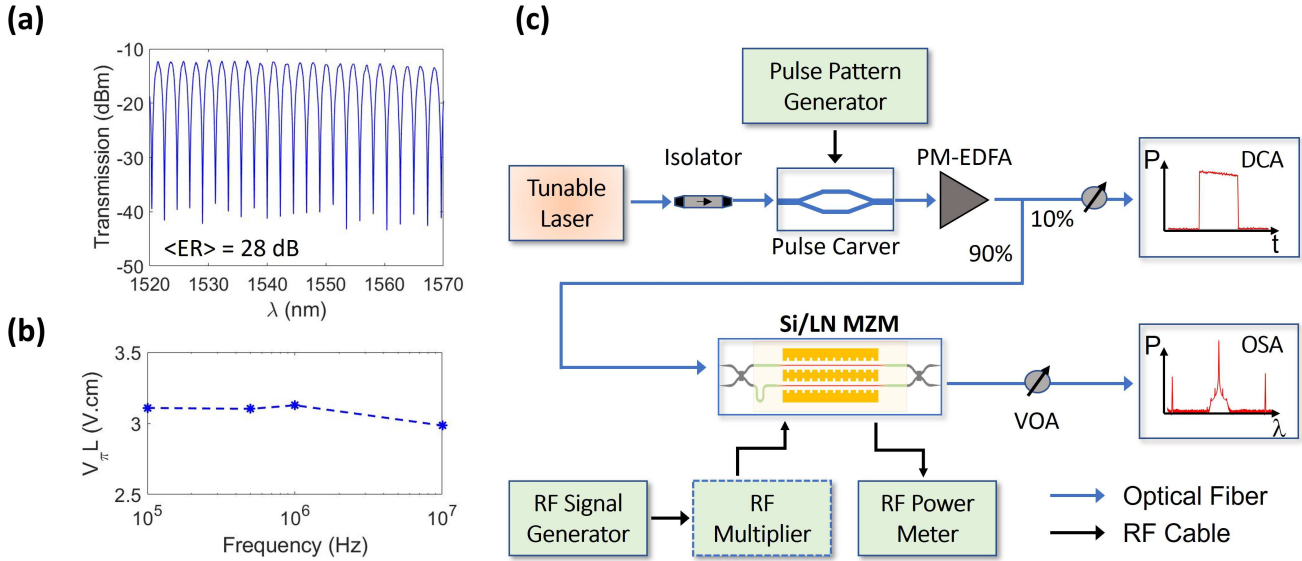}
\caption{\textbf{(a)} The measured optical transmission of the asymmetric hybrid-bonded Si/LN MZM with no modulation signal applied. The asymmetric path-length difference gives an FSR of 2.3 nm, and an average ER of 28 dB over 50 nm. \textbf{(b)} The measured $V_\pi L$ of the hybrid MZM ($L_\mathrm{ps} = 0.5$ cm) as a function of driving trapezoidal signal frequency. \textbf{(c)} A schematic block diagram of the quasi-CW high power electro-optic response measurement setup. Active RF multipliers were used for generating $f >$ 50 GHz.} 
\label{fig4}
\end{figure}

\begin{figure}[ht!]
\includegraphics[width=\textwidth]{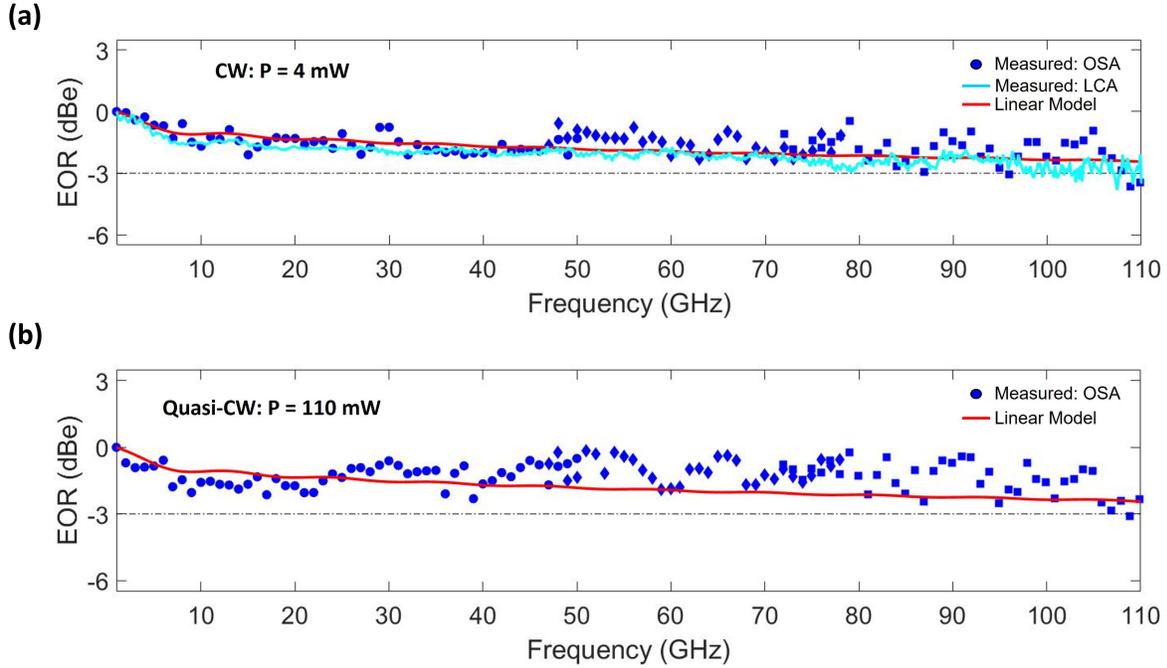}

\caption{The OSA-measured (blue points) EOR of the hybrid bonded Si/LN MZM with $L_\mathrm{ps} = 0.5$ cm using no RF multiplier (circles), a 4x multiplier (diamonds), and a 6x multiplier (squares) resulting in a measured 3-dB bandwidth of 110 GHz for both: \textbf{(a)} Low power, CW optical input with 4 mW of power; and \textbf{(b)} Quasi-CW optical input with peak power of 110 mW. The solid red line is the calculated linear model using Eq. (\ref{eq_EOR}) with no nonlinearities. \textcolor{black}{The cyan line in panel \textbf{(a)} is the EOR of an identically designed Si/LN MZM using a high speed LCA.}}
\label{fig5}
\end{figure}

The asymmetric MZM under test was measured using a tunable CW C-band laser source which resulted in a free spectral range of 2.3 nm, a mean extinction ratio (ER) of 28 dB over 50 nm (maximum of 31 dB at $\lambda = 1560$ nm), and a fiber-to-fiber IL of 12.2 dB as shown in Fig \ref{fig4}(a). The half-wave voltage of the Si/LN modulator with $L_\mathrm{ps} = $ 0.5 cm was measured by overdriving it with a trapezoidal signal and projecting the overshoots of the modulated signal \cite{mere2022modular}, resulting in an average $V_\pi L$ of $3.1$ V\textcolor{black}{.cm} in the range of 0.1 to 10 MHz as shown in Fig. \ref{fig4}(b). Figure \ref{fig4}(c) shows a block diagram of the quasi-CW high power modulation experiment. An external commercial LN modulator (labeled Pulse Carver in Fig. \ref{fig4}(c)) with 30 dB extinction ratio was used to carve the CW laser source into quasi-CW pulses. The pulse carver was driven by a pulse pattern generator, setting voltages with a pulse width of 1 µs and a period of 100 µs. This corresponds to a pulse length greater than 100 m, which is much greater than the length (1.7 cm) of the chip.  A polarization maintaining erbium-doped fiber amplifier (PM-EDFA) was used to amplify the quasi-CW pulses to the Si/LN MZM. Quasi-CW pulses were used to extract maximum gain from the PM-EDFA by reducing the average power and gain saturation. Also, the lower average power reduces the likelihood of damage to the lensed fibers used in the experiments. The amplified pulses were characterized by monitoring the PM-EDFA output using a PM 90-10 splitter, where a sampling oscilloscope (DCA) verified the pulse shape and peak power, while an optical power monitor measured the average power of the pulses as a function of the PM-EDFA current.  Optical attenuators were placed before the DCA and optical spectrum analyzer (OSA) to avoid damage. This device was measured to have an edge coupling loss of 5.2 dB when using lensed tapered fibers with a nominal 2.5 µm mode field diameter; therefore, the on-chip IL of the MZM was \textcolor{black}{1.8} dB (see Methods for more detail). These losses, as well as the fiber, splitter, and connector losses were factored in to calculate the peak power out of the PM-EDFA. The power levels labeled in Fig.~\ref{fig5} and in this discussion refer to the on-chip optical power levels in the feeder waveguide before the start of the MZM section. A peak power level of 110 mW corresponds to an intensity of 460 kW/mm$^2$ and 74 kW/mm$^2$ in the transition and hybrid waveguides, respectively.

The high-power quasi-CW pulses were then coupled to the device and modulated with varying frequency sinusoidal waves using an RF signal generator. A high-resolution OSA was used to detect the generated sidebands and carrier signal as the frequency was swept \cite{Shi2003}. This high-speed modulation measurement was performed in three frequency bands: 1 to 50 GHz, 47 to 78 GHz, and 72 to 110 GHz, where active RF multipliers were used to reach the latter two bands. A CS-5 calibration substrate (GGB Industries, Inc.) and a 110 GHz RF power sensor were used to de-embed the cables and probes used in the experiment and determine the RF power delivered. These calibration measurements were then used to extract the measured EORs of the modulator at relatively low power CW optical input (4 mW), and at input peak powers of 110 mW, as shown in Fig. \ref{fig5}(a) and \ref{fig5}(b), respectively. In both cases, the measured EOR has a 3-dB bandwidth of 110 GHz. The overlapping measured data between the bands was averaged and shifted accordingly to stitch the frequency responses together. \textcolor{black}{Furthermore, the EO $S_{21}$ of an identically designed hybrid bonded MZM on a different chip was independently verified at low CW optical power using a lightwave component analyzer (LCA). The LCA consisted of a Keysight PNA-X network analyzer, frequency extenders, and a 110 GHz photodetector which allowed us to measure modulated signals ranging from 100 MHz to 110 GHz. As can be seen in Fig. \ref{fig5}(a), the OSA-measured response (blue marks) and LCA-measured response (cyan line) agree in both slope and 3-dB bandwidth.} The 3-dB bandwidth is limited from impedance matching. Due to the fixed separation between the Si waveguides of the fabricated SOI chip, the signal electrode to electrode gap spacing is also fixed. This specific MZM was fabricated with an electrode gap of 9 µm, resulting in a transmission line impedance of 42 $\Omega$. If the Si waveguide separation is changed to accommodate a more optimal signal electrode width, then the impedance will be better matched to a 50 $\Omega$ load, and further increase the 3-dB bandwidth of the MZM.

\section*{Discussion}
The red lines in Fig. \ref{fig5}(a) and \ref{fig5}(b) are the calculated EORs using a linear model of the response function of a traveling wave MZM with no intensity dependent nonlinear effects \cite{ghione2009semiconductor}. The theoretical EOR predicts a 3-dB bandwidth above 110 GHz (approximately 156 GHz) for this 0.5 cm long modulator, which is agreement with the measured frequency responses in both the low and high optical power regimes. A large concentration of free carriers overlapping with the optical mode would cause a free-carrier dispersion effect which would decrease the 3-dB bandwidth. However, since the hybrid mode has only 2\% of light in the Si waveguide, there is an insignificant impairment in our device, and the response at 110 mW is effectively the same as that at low (4 mW) power levels. Table \ref{table1} lists performance metrics of recent TFLN-based MZM. The present device is the first (to our knowledge) TFLN MZM that demonstrates greater than 100 GHz modulation of high optical power inputs greater than 100 mW. $\text{FOM}$ (units: dB) in Table \ref{table1} is defined as $\text{FOM} = 20\ \log_{10}(V_\pi) - 20\ \log_{10}(r_d \times P_\text{opt} \times 50\ \Omega)$, where $r_d$ is the detector responsivity, assumed to be 1 A/W. FOM is that portion of the noise figure (NF) expression for an RF photonic link  \cite{Cox2006a} which represents the contribution to the NF from the modulator $V_\pi$ and optical power level. In addition, the laser RIN plays an important role in the overall NF in a practical system, but is not considered here in the FOM for the modulator alone. Lower values of the modulator FOM are preferable because they reduce the minimum achievable NF of the link. This shows the benefits of increasing $P_\mathrm{opt}$ while keeping $V_\pi$ low. The on-chip IL of our hybrid Si/LN device is comparable to that of other TFLN devices. The $V_\pi L$ can be slightly improved by introducing a buffer SiO$_2$ layer between the electrodes and LN film, which will reduce the optical loss due to electrode interaction and allow for a narrower electrode spacing \cite{liu2021wideband}. In comparison to these TFLN MZM devices, a traditional Si depletion-mode MZM made with 250 nm thick Si rib waveguide (90 nm slab height and 650 nm width) will have 75\% of the Poynting energy confined in Si and an $A_\mathrm{eff}$ on the order of 0.2 µm$^2$. Both the linear and nonlinear optical propagation loss coefficients are high, and will limit the high-power handling capabilities of a Si MZM. Furthermore, the modulation speed of carrier depletion MZMs is typically RC-limited, with reported 3-dB electro-optic bandwidths $<$ 60 GHz \cite{watts2010low,dong2012high,derose2012high,xiao2013high,li2018silicon}.

\newcolumntype{R}{>{\raggedleft\arraybackslash}p{1.5cm}}
\begin{table}[ht]
\begin{center}
    \renewcommand{\arraystretch}{1.3}
    \caption{Comparison of recent TFLN-based MZM Performance Metrics}
    \centering
    \label{table1}
    \begin{tabular}{c c c c c c c c}
    \toprule%
    Device & 3-dB BW (GHz)& $V_\pi L$ (V.cm) & IL (dB) & $L_\mathrm{ps}$ (cm) & $P_\mathrm{opt}$ (mW) & $\text{FOM}$(dB)\\
    \midrule
    Etched LNOI\cite{wang2018integrated} & 100 & 2.2 & 0.5 & 0.5 & 1 & 38.9 \\
    Etched LNOI\cite{yang2022monolithic} & $>110$ & 2.4 & 18 (fiber-to-fiber) & 0.5 & 1 & 39.6 \\
    Etched LNOI\cite{shams2022electrically} & 50 & 2.2 & NR & 0.5 & 25 & 10.9 \\
    Etched LNOI \cite{liu2021wideband} & >67 & 1.7 & 17 (fiber-to-fiber) & 0.5 & NR & - \\
    Etched LNOI \cite{chen2022high} & >67 & 2.2 & 0.2 & 1 & 0.03 & 63.3\\
    \textcolor{black}{Bonded (BCB-assisted) Si/LN \cite{wang2022silicon}} & \textcolor{black}{>70} & \textcolor{black}{2.1} & \textcolor{black}{3} & \textcolor{black}{1.25} & \textcolor{black}{NR} & \textcolor{black}{-}\\
    Polished LNOI\cite{wu2022high} & >50 & 2.2 & 0.6 & 0.7 & NR & - \\
    \textcolor{black}{Bonded (BCB-assisted) Si + Etched LN} \cite{he2019high} & >70 & 2.5 & 2.5 & 0.5 & NR & -\\
    \textcolor{black}{Bonded (Direct)} SiN/LN\cite{Boynton2020a} & >50 & 6.7 & 13 (fiber-to-fiber) & 0.5 & 4 & 36.5\\
    \textcolor{black}{Bonded (Direct) Si/LN - } This Work & 110 & 3.1 &  \textcolor{black}{1.8} & 0.5 & 110 & 1.04 \\
    \bottomrule
\end{tabular}
\end{center}
\footnotesize{NR: Not Reported. IL (dB): on-chip Insertion Loss in dB; fiber-to-fiber Insertion Loss in dB is reported where noted. $\text{FOM}$(dB) is defined as $20\ \log_{10}(V_\pi)-20\ \log_{10}(P_\text{opt}r_{d}R_{s})$, where $r_d = 1$ A/W is the photodetector responsivity (assumed) and $R_s = 50$ $\Omega$ is the source resistance. Entries are left blank where $P_\mathrm{opt}$ was not reported.}
\end{table}
Our experimental observations are consistent with the theoretical predictions shown in Fig. \ref{fig2}(b). Using Eq. (\ref{eqn_VpiL}), the effect of TPA induced carriers on the $V_\pi L$ through FCD [Fig. \ref{fig2}(b)] is insignificant at 110 mW, and would contribute less than 1\% change in modulation efficiency assuming thermo-optic and photorefractive effects in the TFLN layer are negligible.  The change in IL due to carrier effects in the Si regions are predicted to be under 1 dB as shown by Fig. \ref{fig2}(a). Practically, such small changes  are indistinguishable from other losses in this study, such as repeatability of fiber coupling in these bare-die test chips. Fiber-pigtailing the device will allow for easier testing. Although these tests were performed using long quasi-CW pulses, we expect similar performance, when packaged, under CW optical excitation as well. 

As we scale to longer phase shifters ($L_\mathrm{ps} > 1$ cm) to reduce the voltage, the walk-off between the RF and optical phases can occur from smaller velocity differences. Therefore, the changes caused by higher optical power can have a larger effect on the 3-dB bandwidth. However, our calculations show that the majority of the impairments will occur in the mode transition region rather than the phase-shifter region. The theoretical minimum length for a low-loss adiabatic taper between the transition mode and hybrid mode in this design is about 100 µm \cite{fu2014efficient}. In this batch of test chips, the transition waveguides were intentionally made longer for fabrication ease, and therefore the nonlinear losses will inevitably be higher. The feeder waveguide and taper lengths could be reduced, which in turn will minimize the TPA-induced FCA of the transition section. If the combined input and output transition sections were only 100 µm long instead of 1.4 mm long as in these chips, then the additional insertion loss due to TPA would incur a 1 dB penalty only for optical powers greater than 0.6 W. For higher power operation, other materials with lower multiphoton absorption effects can be used. As an example, silicon nitride is another CMOS-compatible material, does not experience TPA in the telecom wavelength regime, and has been shown to be suitable for hybrid bonding with unpatterned LNOI \cite{Boynton2020a}.

In summary, we have demonstrated that hybrid TFLN MZMs, which use an unetched thin-film of LN bonded to planarized crystalline Si waveguides, can withstand high optical input powers of 110 mW while maintaining high speed modulation bandwidths of 110 GHz. A high EO bandwidth was achieved by using an inductively loaded SWE design to  achieve velocity matching between the RF and optical waves. The hybrid Si/LN device design was optimized to achieve a high modulation efficiency ($V_\pi L = 3.1$ V.cm) while minimizing the nonlinear intensity dependent effects of Si. The on-chip IL of the hybrid MZM structure including the MMI couplers, transitions into and out of the bonded region and electrode interaction loss was measured to be \textcolor{black}{1.8} dB. High optical power operation was tested by amplifying quasi-CW pulses with a pulse width of 1 µs and at power levels of up to  110 mW, we did not observe degradation of the EO bandwidth compared to measurements made at a conventional power level (4 mW). Compared to measurements made at an input power of 6 dBm, the additional nonlinear insertion loss was less than 1 dB at 20.4 dBm optical power. Our work demonstrates the integration of Mach-Zehnder modulators with high bandwidth and power handling capabilities well beyond the capabilities of traditional Si photonics.

\section*{Methods} 

\subsection*{Insertion Loss Characterization}
\begin{figure}[ht!]
\includegraphics[width=\textwidth]{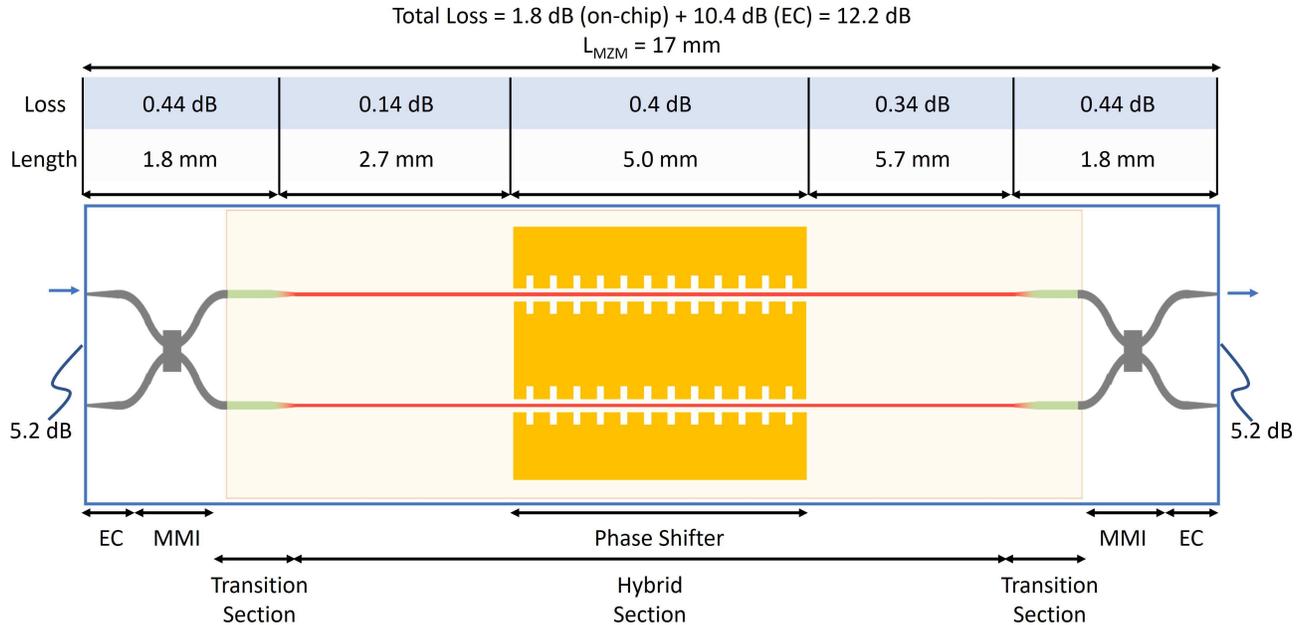}
\caption{A (not to scale) schematic of the hybrid bonded Si/LN MZM showing the total measured insertion loss and the loss attributed to the edge coupler (EC), transition, hybrid, and phase shifter sections.}
\label{figappIL}
\end{figure}

To determine the low power (1 mW) IL of the hybrid bonded Si/LN MZM, a CW tunable laser was coupled to the device using lensed tapered fibers (2.5 µm nominal mode field diameter). As shown in Fig. \ref{figappIL}, the hybrid bonded Si/LN MZM consists of edge couplers (ECs), MMIs, wide Si transition sections and a narrow Si hybrid section. The propagation loss of the transition waveguide section is 0.8 dB/cm, measured using the cutback method \textcolor{black}{\cite{wang2022integrated}}; whereas the propagation loss of the hybrid section was reported in an earlier work as 0.6 dB/cm \cite{weigel2018high}. To determine the IL of the 3-dB 2x2 MMI splitters, adjacent test MMIs of the same design was measured, resulting in a loss of 0.2 dB per splitter. The 1 dB bandwidth of the MMI is over 50 nm with an imbalance between the output ports less than 0.3 dB. The IL of the phase shifter section was measured by subtracting the transmission spectra of the test device and an identical reference device, but without electrodes. Therefore, the IL of the $L_\mathrm{ps} = 5$ mm long phase shifter is 0.4 dB. From these measurements, the individual component losses of the MZM were subtracted from the total transmission, resulting in an edge coupling loss of 5.2 dB/facet and an on-chip loss of \textcolor{black}{1.8} dB (assuming 0.1 dB of loss per LN edge transition) as shown in Fig. \ref{figappIL}.

\section*{Data availability.} Data underlying the results presented in this paper are not publicly available at this time but may be obtained from the corresponding author upon reasonable request.


\bibliography{References_110GHz_110mW_SiLN_noDOI}

\section*{Additional Information}
\subsection*{Competing Interests}
The authors declare no competing interests.

\end{document}